\documentclass[a4paper,amsmath,amssymb,
reprint,%
]{revtex4-1}
\usepackage{graphicx}% Include figure files
\usepackage{dcolumn}% Align table columns on decimal point
\usepackage{bm}% bold math
\usepackage{pifont} % for circled numbers
\usepackage[utf8x]{inputenc}
\usepackage{xcolor}
\usepackage{siunitx}
\usepackage{csquotes}
\usepackage{float}
\usepackage{subfigure}

\newcommand\revision[1]{\textcolor{black}{#1}}

\newcolumntype{M}[1]{>{\centering\arraybackslash}m{#1}}

\begin{document}

\preprint{}

\title{Intermittency of quantum turbulence with superfluid fractions from 0\% to 96\% % in a high Reynolds number wind-tunnel
}
%\thanks{Footnote to title of article.}

%\affiliation{}

\author{E. Rusaouen$^{1}$ , B. Chabaud$^{1}$, J. Salort$^2$, P.-E Roche$^{1}$}
\affiliation{$^1$ Univ. Grenoble Alpes, CNRS, Grenoble INP, Institut Néel, 38000 Grenoble, France\\
$^2$Univ Lyon, Ens de Lyon, Univ Claude Bernard, CNRS, Laboratoire de Physique, F-69342 Lyon, France.\\
%$*$footnote : present address: LEGI, CNRS/UGA/Grenoble INP, F-38000 Grenoble, France.
}

\date{\today}

\begin{abstract}

The intermittency of turbulent superfluid helium is explored systematically in a steady wake flow from 1.28 K up to $T>2.18K$ using a local anemometer. This temperature range spans relative densities of superfluid from 96\% down to 0\%, allowing to test numerical predictions of enhancement or depletion of intermittency at intermediate superfluid fractions. Using the so-called extended self-similarity method, scaling exponents of structure functions have been calculated. No evidence of temperature dependence is found on these scaling exponents in the upper part of the inertial cascade, where turbulence is well developed and fully resolved by the probe. This result supports the picture of a profound analogy between classical and quantum turbulence in their inertial range, including the violation of self-similarities associated with inertial-range intermittency.

\end{abstract}



\maketitle

\section{Motivation and state-of-the-art}

\subsection{Introduction}
 
When liquid $^4$He is cooled below $T_\lambda \simeq 2.18 K$, it experiences a phase transition and enters a superfluid state, called He-II.
The hydrodynamics of superfluids has fascinated physicists and engineers since the late 1930s, in particular for their ability to flow without experiencing any viscosity, and for the quantification of vorticity, discovered a decade later \cite{DonnellyLivreVortices,VanSciverLivre2012}. Due to their exotic properties, He-II and other quantum fluids have also attracted interest from the classical turbulence community, as it allows to tackle some open problems using a fluid with unique dissipative and vorticity properties \cite{BarenghiSkrbekSreenivasan_IntroPNAS2014}. 
The so-called \textit{quantum turbulence} of mechanically-stirred He-II was found to share many features with classical turbulence \cite{spectra:PNAS2014}, in particular in the so-called \textit{inertial range} of scales, where the kinetic energy continuously cascades from larger to smaller eddies \cite{salortTOUPIE}, resulting in a Kolmogorov-Obhukov-like $k^{-5/3}$ velocity power spectrum ($k$ is the wavenumber). 
The present study explores the phenomenon of intermittency in this inertial range, an effect associated with a violation of self-similarity of velocity fluctuations, which is still actively studied in classical turbulence \cite{uriel1995turbulence,Sreenivasan:1997p84,tsinober2013essence,benzi2015homogeneous}.

Using Landau and Tisza two-fluid model, He-II hydrodynamics can be described by two interpenetrating fluids in mutual interaction : one inviscid \textit{superfluid}  of density $\rho_s$ and one \textit{normal fluid} of viscosity $\mu$ and density $\rho_n=\rho - \rho_s$ (where $\rho$ is the density of He-II) \cite{DonnellyLivreVortices,VanSciverLivre2012}. By changing the temperature between $T_\lambda$ and 0K, the superfluid fraction $\rho_s/ \rho$
 can be arbitrarily chosen between 0\% and 100\%. This temperature dependence is a key property of the present study: it allows to explore intermittency from the Navier-Stokes case ($T>T_\lambda$ and $\rho_s / \rho=0 \%$),  down to a nearly pure superfluid (here  $\rho_s / \rho \simeq 96 \%$). The universality of intermittency can therefore be tested versus a continuous change of fluid properties.

\subsection{Contradictory numerical predictions}

\begin{center}
\begin{table}
\begin{tabular}{M{2cm} | M{2cm} | M{2cm} | M{2cm}}
Reference & Approach & Superfluid fraction & Intermittency exponents \tabularnewline
&& $\rho_s/\rho$ & ($\zeta_{p \geqslant 3}$ )  \tabularnewline
 \hline
 \hline
Maurer and Tabeling \cite{Tabeling} & Experiment & $92\%$ & Consistent with classical  \tabularnewline		\hline
Salort \textit{et al} \cite{salortInter} & Experiment &  $0\%$ and $85\%$ & Consistent with classical  \tabularnewline
 & DNS simulations (based on HVBK) & $9\%$ and $98\%$ & Consistent with classical  \tabularnewline
\hline
Boué \textit{et al} \cite{LvovInter} & Shell-model simulations& $\textit{ $\sim20-90\%$}$ &\textit{More intermittent} \tabularnewline
&  (based on HVBK)& $\lesssim 20\%$ or $\gtrsim90\%$ & Consistent with classical  \tabularnewline
\hline
Shukla and Pandhit \cite{Shukla} &  Shell-model simulations& $\textit{$\sim10-80\%$}$ & \textit{Less intermittent} \tabularnewline
& (based on HVBK) & $\lesssim 40\%$ or $\gtrsim65\%$ & Consistent with classical  \tabularnewline
\hline
Bakhtaoui and Merahi \cite{Bakhtaoui:2014} & LES simulations &  $\textit{84\%}$ & \textit{More intermittent} \tabularnewline
 & (based on HVBK) & $23\%$ and $98\%$ & Consistent with classical  \tabularnewline
\hline
Krstulovic \cite{Krstulovic:PRE2016} & Gross-Pitaevskii simulation &  $\textit{100\%}$ & \textit{More intermittent} \tabularnewline
\hline
Rusaouen \textit{et al} \cite{Rusaouen:parietalEPL2017} & Experiment & $0\%$, $19\%$ and $81\%$ & Consistent with classical  \tabularnewline
\hline
Rusaouen \textit{et al} (present study) & Experiment & $0\%$, $11.3\%$, $51\%$, $63\%$, $85.8\%$ and $95.7\%$ & Consistent with classical \tabularnewline
\hline
\end{tabular}
\caption{\label{Review}Experimental and numerical studies of quantum turbulence intermittency. The statements ``more'' or ``less'' intermittent are based on structure functions of order larger than two (eg. as shown on figure \ref{ExponentP-NS}). The second order structure function can suggest an opposite trend.}
\end{table}
\end{center}

For convenience, Table I summarizes the literature review presented in the following paragraph.

The first experimental studies of intermittency in superfluid were published in 1998 and 2011 \cite{Tabeling,salortInter}. They focused on the low temperature regime with superfluid fractions $\rho_s/\rho=92\%$ and $85\%$ (respectively 1.4K and 1.56K). Both experiments reported no difference with the intermittency of classical fluids.

In 2011, some direct numerical simulations (DNS) based on the so-called HVBK continuous model \cite{DonnellyLivreVortices,VanSciverLivre2012} were also reported in reference \cite{salortInter}. In the HVBK model, the quantized nature of the superfluid vorticity is coarse-grained into a continuous field, which allows to describe the fluid using an Euler equation (for the superfluid) and a Navier-Stokes (for the normal fluid) coupled by a mutual friction term. In the DNS study mentioned above,  both the low and high temperature regimes were explored, with superfluid fractions of 98\% and 9\% respectively. Again no difference was found with the classical fluid intermittency. 

In 2013,  Boué et al. \cite{LvovInter} reported numerical simulations using a shell-model \cite{biferale2003shell} of the HVBK dynamics. In the low and high temperature limits, they found the same results as the previous studies. 
But they also reported a significant  \textit{enhancement} of intermittency at intermediate temperature, corresponding to the window $\rho_s/ \rho \simeq 20-90\%$ (yet the exponent of the second order structure function reaches values corresponding to an absence of intermittency)  \cite{LvovInter}. 
In 2016, numerical studies by Shukla et Pandit \cite{Shukla} using a different variant of shell model (respectively Sabra version and a GOY variant) agreed on the low and high temperature limits but reported opposite results in the intermediate window
 with a significant \textit{reduction} or \textit{absence} of intermittency (the second order structure function exhibits a more complex behavior). No  experimental data was available for comparison in this intermediate temperature range.

In 2014, some Large Eddy Simulations (LES) of the HVBK model were reported for $\rho_s / \rho=98\%$, 84\% and 23\% by Bakhtaoui and Merahi \cite{Bakhtaoui:2014}.
The authors report a significant difference of intermittent behavior at their intermediate temperature ($T=1.6$K,  $\rho_s / \rho=84\%$) compared to their lowest and highest temperature cases, and they interpret it as a signature of intermittency \textit{enhancement}.

Adding to the apparent puzzle, another 2016 study explored quantum-fluid intermittency at zero temperature ($\rho_s / \rho=100\%$) using Gross-Pitaevskii equations, and concluded on intermittency \textit{enhancement} \cite{Krstulovic:PRE2016}. Once again, no experimental data on intermittency is available today in this zero temperature case where the normal fluid fraction is null.

Finally, a 2017 experimental study in a highly turbulent Von Karman cell ($R_\lambda \sim 10000$) took a different perspective by analyzing the intermittent statistics of coherent structures for $\rho_s / \rho= 0\%$, 19\% and $81\% (\pm2\%)$. No temperature dependence was found \cite{Rusaouen:parietalEPL2017}, like in the previous experimental studies.

As a side note, we can mention for completeness two on-going studies have been reported by Emil Varga and Victor L'vov in the Quantum Turbulence workshop held in Tallahassee in April 2017 : one experimental work on transverse structure functions performed in Tallahassee and numerical simulations performed by DNS in collaboration between groups from Rehovot and Rome. \revision{Some DNS simulations using the HVBK model have also been performed lately in Rahul Pandit's group (private communication)}.

The puzzle of these contradictory numerical results and the lack of experimental data at intermediate temperatures motivated the present systematic experiment.

\subsection{Methodology}

Our experimental aim is a high-resolution assessment of the temperature-dependence of intermittency in the inertial range of turbulent helium from its classical state ($T>T_{\lambda}$) to its superfluid one, down to temperature corresponding to a superfluid fraction of $\rho_s /\rho=96 \%$.

An accurate determination of intermittency  is only possible under several conditions. One condition is a good convergence of velocity statistics, which led to the choice of a steady flow rather than an unsteady one. A second condition is %XX
to have a sufficiently large inertial range \cite{Kahal}
and a third condition is to resolve its velocity fluctuations : here we cover more than 1.5 decades of frequencies, as illustrated on Fig. \ref{spectra}.

After considering different types of flows, such as grid and Von Karman flow, we chose to study the turbulence in the wake of a disc. Furthermore, the flow was confined in a pipe to preserve a well-defined mean direction. Although wake turbulence is not isotropic nor homogeneous, it appeared as a good compromise to meet the requirements listed above and to explore the temperature dependence of intermittency in a well-defined developed turbulent flow.
Wakes of discs have been widely studied in classical turbulence (e.g. \cite{Carmody1,Cannon,Johansson,Johansson2}), and even in superfluid helium for one of  intermittency studies previously mentioned \cite{salortInter}. Thus, the existing literature allows to size the experiment (see Section II-a) and probes (see Section II-b) in order to generate a well-defined turbulent flow (see Section III-a).

Compared to the previous experimental studies \cite{Tabeling,salortInter}, the flow temperature is varied systematically and over a broader range. Reference measurements are performed above and below the superfluid transition in conditions as similar as possible, to allow one-to-one comparison.
To allow a direct comparison with the previous works cited above, intermittency is quantified by the exponents of the velocity structure functions, as discussed later (Section III-b).

\section{The TOUPIE experiment}

\subsection{Experimental setup}

\begin{figure}
\begin{center}
\subfigure[General view]{\label{sketchTOUPIEa}\includegraphics[height=12.5cm]{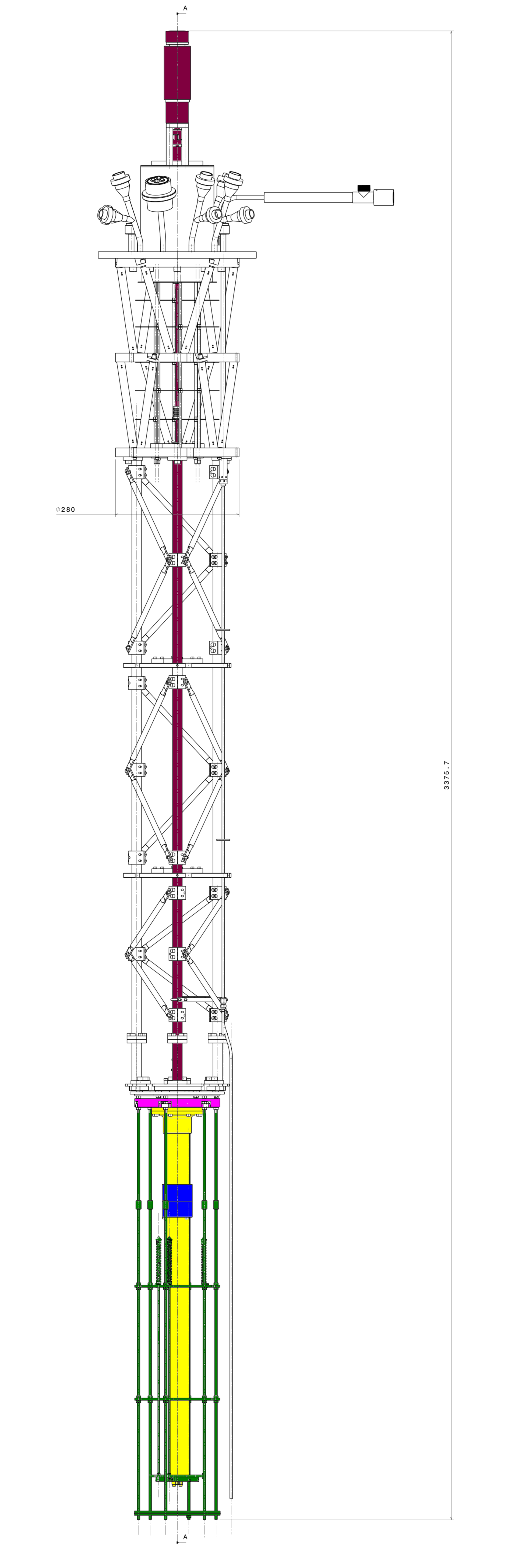}}
\subfigure[Zoom on the wind tunnel part.]{\label{sketchTOUPIEb}\includegraphics[height=12.5cm]{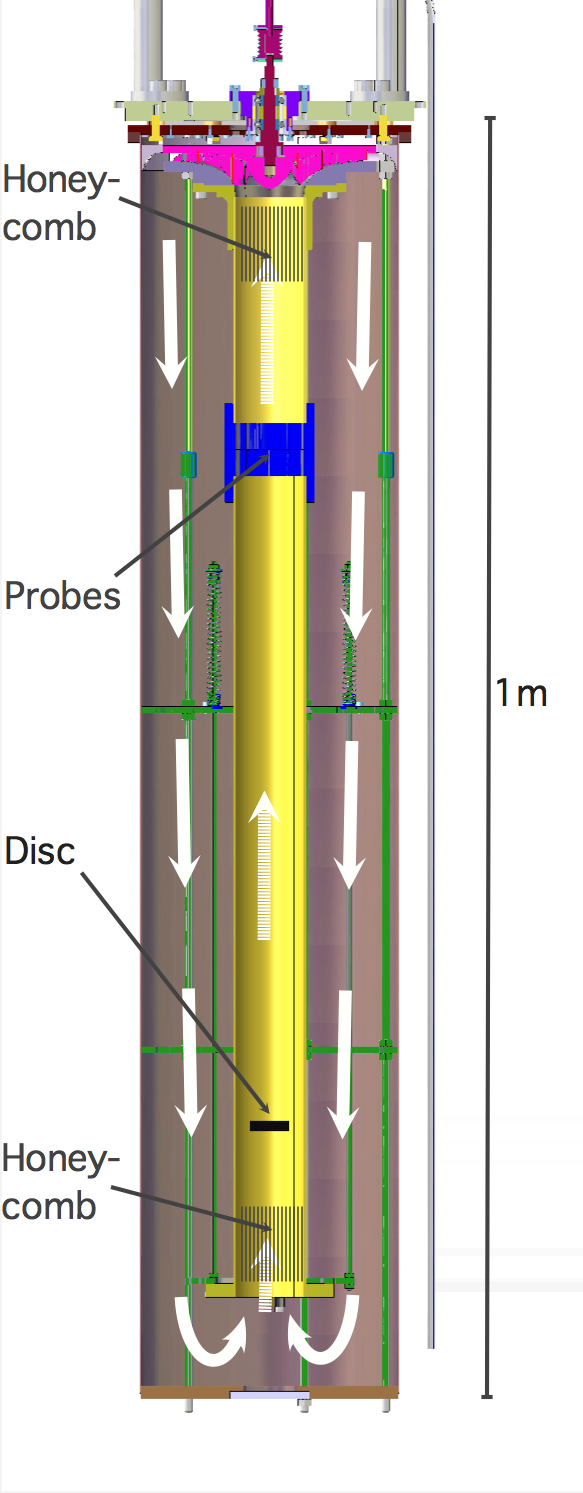}}
\end{center}
\caption{The TOUPIE experiment : yellow part is the inner wind tunnel, blue part is the instrumentation support, green tends for the decoupling springs and the wake-generator disc is in black. The room temperature motor and drive shaft are schematized in violet and the propeller is in pink color. (image © Ph.R.)}
\end{figure}

The TOUPIE liquid helium wind-tunnel, previously described in \cite{salortTOUPIE}, has been upgraded and adapted to the requirements of the experiment.
It consists in a 1-m-long wind-tunnel, mounted at the bottom end of a cryogenic insert exceeding 2 m in length (see figure \ref{sketchTOUPIEa}). Such a long insert allows an hydrostatic pressurization of the wind tunnel under a column of liquid helium exceeding $h=1$\,m in height, which prevents cavitation up to flow velocities exceeding $\sqrt{2 g h} \simeq 4.4\,$m/s.  The insert is designed to provide high stiffness to the experiment thanks to the truss structure visible on the general view in figure \ref{sketchTOUPIEa}.

The wind-tunnel itself has a coaxial cylindrical geometry : the test section is within the inner cylinder while the return channel is between the inner and outer cylinders (see fig.
\ref{sketchTOUPIEb}). The 19.5-cm-diameter outer cylinder is made of a thin Cu sheet for efficient energy exchange with the surrounding cooling helium bath, while the inner cylinder is a  \SI{80}{cm}-long and \SI{5.1}{cm}-internal-diameter cardboard tube (in yellow on the sketch). Cardboard is chosen to reduce the propagation of vibrations. It is partly decoupled from the rest of the structure by three springs (in green on the sketch). The cardboard tube (from the roll of a poster-printer paper) is interrupted by a massive brass ring at the location where the probes are mounted (in blue on the sketch). The spring stiffness is chosen just as large as required to support  slightly more the weight of the brass ring and tube ($\simeq 1.5$\,kg). This allows to benefit from the low pass filter of this mass-spring mechanical resonator.

Reminiscent of the design of so-called \textit{étoile flow conditionner}, six flow-guides made of Kapton sheets prevent helicoidal motion of the flow along the return section
. In the same spirit, two honeycombs are inserted at the entrance of the inner pipe and at its end, right upstream the propeller. Both honeycombs exhibit the same cells density: \si{10} {cells/cm$^2$} and respective length of \SI{5}{cm} (input of the wind-tunnel) and \SI{2}{cm} (output of the wind-tunnel). They main purpose is to straighten the flow, remove swirl and lower to turbulence intensity \cite{mehta_bradshaw_1979}.
 
Flow instabilities  sustain an acoustic standing wave settling in the helium, between the top and bottom walls of the wind-tunnel. To reduce its  impact on the Pitot tube measurement (see next sub-section), the probe-holding brass ring was initially located at mid-height in the tunnel, where 1$^{st}$ mode of the standing wave has a pressure node. The improvement on the acoustic pollution captured by the Pitot tube was found marginal and this probe-positioning constrain was  abandoned.
 
The fluid is set into motion by a centrifugal pump optimized to reach a mass flow of \SI{130}{g/s} of liquid helium. A drive shaft connects the pump to a motor at ambiant temperature. 
Special attention was paid to the stainless steel ball bearing located at the bottom of the shaft since past experiments have shown that it can be a source of vibrations in the few hundreds of Hz range. For cost reasons, we use standard stainless steal bearings, cleaned in a solvant to remove the lubricant oil which would freeze at low temperature. Unsurprisingly, these oil-free bearings aged more rapidly, even when dry lubricants are added, which result in more vibrations. As a consequence, a new bearing is mounted before each cool-down of the wind-tunnel. To spoil the acoustic impedance matching coupling between the stainless steal bearing and the stainless steal plate on which it is fixed, a fiber-glass-reinforced epoxy cage in inserted in-between (in purple on figure \ref{sketchTOUPIEb}).

Rotation of the shaft ($\Omega$ in \SI{}{\hertz}) is measured using a dynamo, and is proportional to the velocity of the fluid $V$ in the test-section, up to small corrections due a reduced efficiency of the pump at the lowest rotation frequencies.
Unfortunately, the proportionality coefficient -around few tens of Hz/(m.s$^{-1})$- was not measured accurately due to a technical problem. So velocity is kept in arbitrary units of propeller rotation.

A disc of diameter $d = \SI{25.5}{mm}$,  \SI{3.7}{\milli \meter} thickness with sharp edges generates a turbulent wake in the test section. For the maximum He mass flow of \SI{130}{g/s}, and a density of $\rho=145$kg/m$^3$, the wind-tunnel has been designed to reach a maximum mean velocity is $\left< V \right> \approx \SI{0.5}{m/s} $ with the present pipe section. In this work, the rotating velocity is half the maximum one, corresponding to a disc Reynolds number 
\begin{equation}
Re_d = {{\left<V \right> d} \over {\nu}} \simeq 3.10^5.
\end{equation}
for a kinematic viscosity $\nu=\mu/\rho$ taken at \SI{2.32}{\kelvin}. A rough estimation of the Taylor microscale Reynolds number at the location of the probe can be made assuming an integral scale of $d/2$ and a turbulence intensity of 7\% (as measured), $R_\lambda = \sqrt{{15 \times 0.07 V d/2 \nu}} \sim 400$

Temperature is decreased below \SI{4.2}{\kelvin} by pumping the helium bath with a roots group (Leybold model SV300 and WS2001). With such a flow, temperature can be as low as \SI{1.28}{\kelvin} at the largest Reynolds numbers. This corresponds to a superfluid fraction of nearly 96\% (see table \ref{Table}). Experiments have been performed at seven different temperatures.

\begin{center}
\begin{table}
\begin{tabular}{c | c | c | c}
Temperature & Pressure & Superfluid & Kinematic\\
& & fraction& viscosity\\
$T$ [\SI{}{\kelvin}] & $P$ [\SI{}{mBar}]& $\rho_s/\rho$ & $\mu/\rho$ [\SI{}{m^2/s}]\\
 \hline
 \hline
2.32 & 70 & 0 & $ 1.99\cdot10^{-8}$\\			%2.18\cdot10^{-8}
2.15 & 47.3 &  $11.3\%$ & $1.52\cdot10^{-8}$\\
2.10 & 41.3 & 25\% & $1.28\cdot10^{-8}$\\
1.95 & 26.9 & 51\% & $9.57\cdot10^{-9}$\\
1.85 & 19.6 & 63\% & $9.00\cdot10^{-9}$\\
1.55 & 6.08 & 85.8\% & $9.50 \cdot10^{-9}$\\
1.28 & 1.42 & 95.7\% & $1.24\cdot10^{-8}$
\end{tabular}
\caption{\label{Table}Flows main characteristics. Temperature is computed from the measured pressure of the saturated liquid. Densities and kinematic viscosity are computed using the temperature and pressure corrected from the hydrostatic pressure.}
\end{table}
\end{center}

 \subsection{Instrumentation}
 
 \paragraph*{Probes:}
 
Two probes, a micro-cantilever anemometer and a miniature total head-pressure probe (later referred to as ``Pitot tube'')  are inserted in the test section.

\begin{figure}
\begin{center}
\subfigure[Sketch of a cantilever probe]{\label{Probes1}\includegraphics[width=6.5cm]{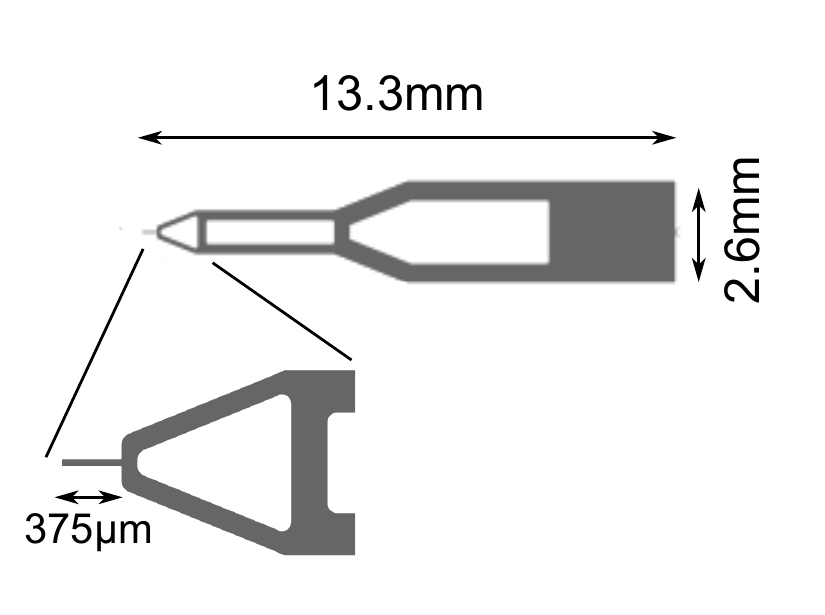}}
\subfigure[Electronic microscope image of the cantilever]{\label{Probes1sem}\includegraphics[width=6cm]{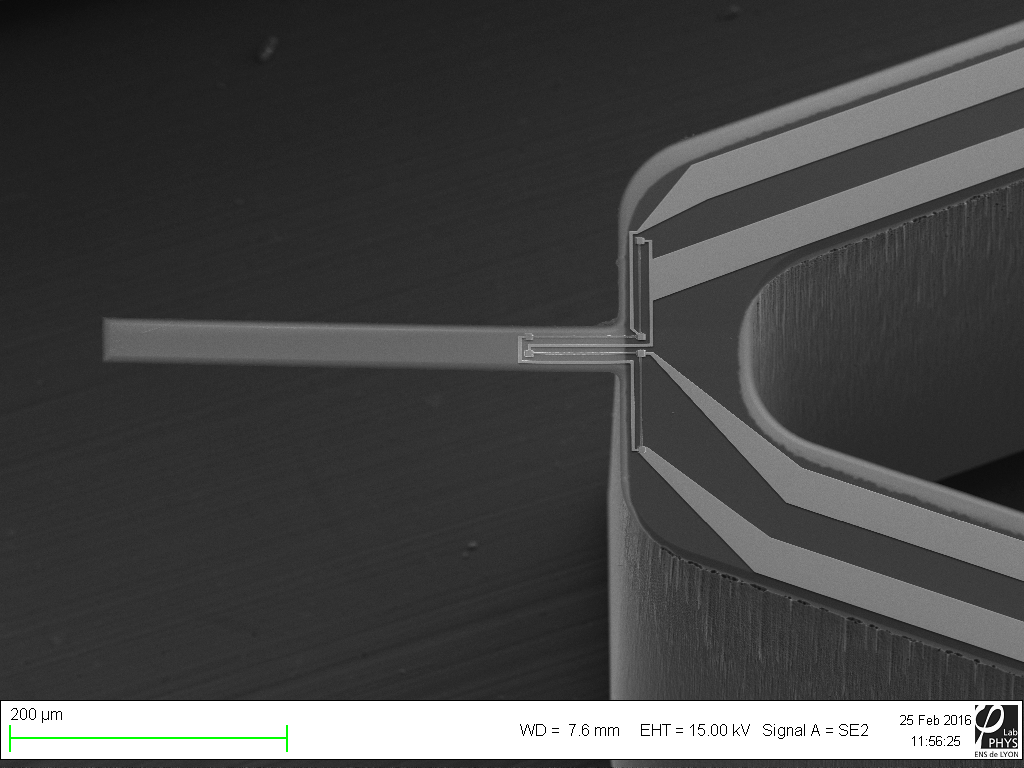}}
\subfigure[View of the probes]{\label{Probes2}\includegraphics[width=6cm]{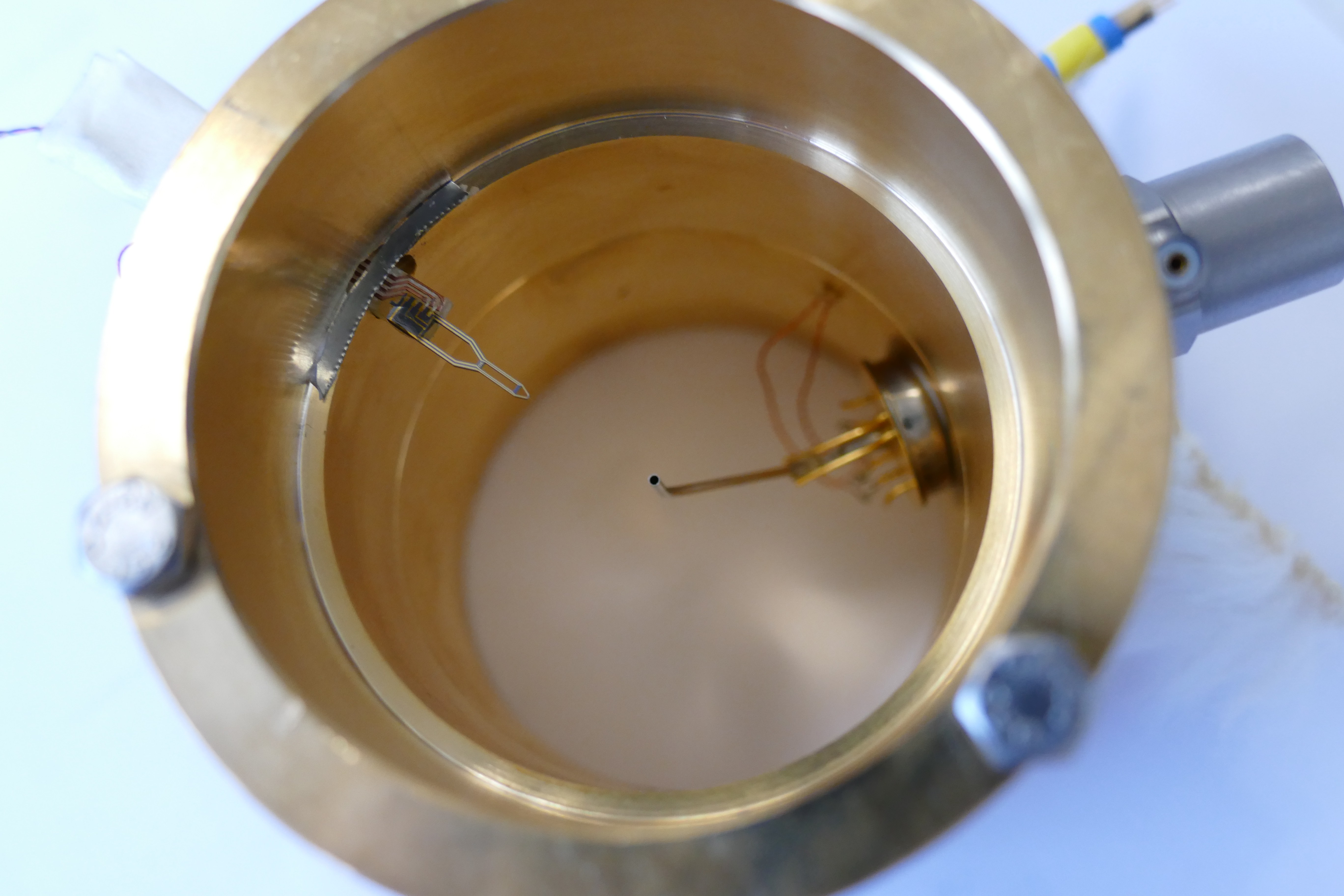}}
\end{center}
\caption{Two probes were inserted in the wind tunnel : a Pitot tube and micro-machined cantilever anemometer}
\end{figure}

The micro-machined cantilever is sketched in figure \ref{Probes1}, above an electron microscope image. It consists in a rectangular beam, \SI{375}{\micro \meter} long, \SI{32}{\micro \meter} large and \SI{1.2}{\micro \meter} thick, made of silicon oxide which is deflected by the incident flow. Both probes are sensitive to the local dynamic pressure $\frac{1}{2}\rho u^2$. The cantilever beam, its supporting structure and its built-in resistive strain gauge are machined using micro-system technics in clean room.
 Details about the manufacturing process can be found in \cite{salortMEMS1,salortMEMS2}. % XXX
  The first resonance frequency of the cantilever immerged in liquid helium is estimated to be around 5 kHz \cite{Sader_JApplPhys98,salortMEMS2}, which is above the range of frequency of interest in the present study (typ. DC-1 kHz).

The Pitot tube is built with a capillary tube of internal diameter \SI{0.8}{mm} and \SI{34}{mm} long, parallel to the mean flow at one end and closed with a micro-machined differential piezo-resistive pressure transducer at the other end. The Helmholtz resonance of this probe at ambiant temperature is close to \SI{1500}{\hertz}, leading to a \SI{500}{\hertz} resonance at  \SI{2}{\kelvin}, due to the 3 ratio between sound velocity in atmospheric air and in liquid helium. Unfortunately, acoustic perturbations have polluted the signal and significantly reduced the exploitable frequency range down to 70 Hz typically. As a consequence, we only use the Pitot tube to validate the mean response of the cantilever probe and the efficiency of the centrifugal pump, using the well-known quadratic response of Pitot tubes versus velocity.
 
 Both miniature Pitot tubes and micro-cantilevers have been previously validated for anemometry of the longitudinal velocity component in wind-tunnels, above and below the superfluid transition \cite{Tabeling,salortMEMS1,salortTSF,salortTOUPIE}. In He-II, both anemometers are sensitive to the barycentric velocity of the superfluid and normal fluid $V= V_s \rho_s/\rho + V_n \rho_n/\rho$ (with obvious notation). But at the inertial scale resolved by the probes, the two fluids are known to be locked together in this temperature range \cite{Roche2fluidCascade:EPL2009}, and the probes are thus sensing the common velocity: $V \simeq  V_s  \simeq  V_n$. \\

 \paragraph*{Position in the flow:}
 
 Reference \cite{Carmody1} shows that wake turbulence downstream a disc becomes fully developed (ie: self-similar) at 15 disc diameters, for an unconfined flow with $Re_d \simeq 7.10^4$. 
In our experiment, $Re_d$ is half a decade larger and the disc of diameter $d=2.5$cm is  confined in a tube of diameter 5.1cm, which obviously results in different streamwise flow properties. To our knowledge, no study of this particular issue in the wake of a disc and at such large Reynolds number exists. Thus, we have chosen to place the probes 20 disc diameters downstream the disc of diameter $d$ and we don't expect turbulence to be fully developed down to the smallest scales of the inertial range

The Pitot tube is located on the axis of the test section whereas the cantilever is $d/6$ aside the axis, see picture \ref{Probes2}. 

 \subsection{Measurement protocol}

\begin{figure}
\begin{center}
\includegraphics[width=8.5cm]{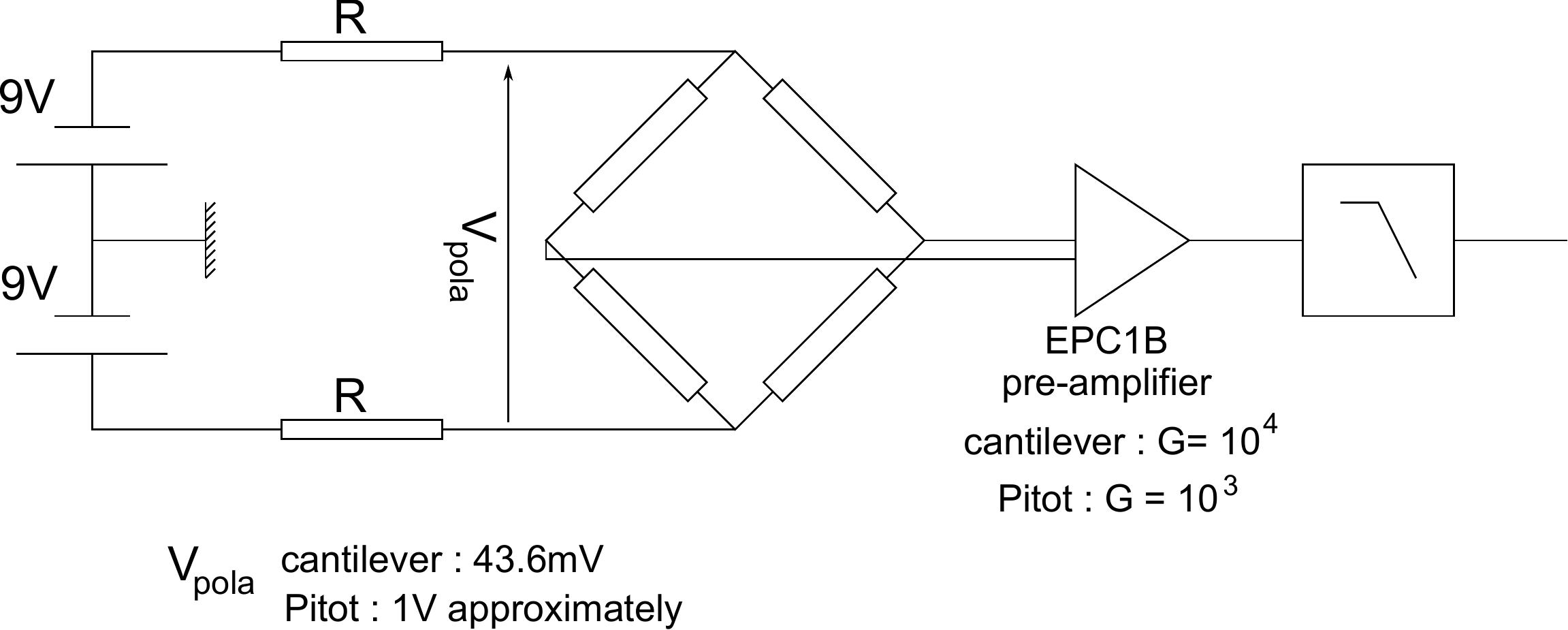}
\end{center}
\caption{\label{sketchElecFluct}Electrical circuit used for fluctuations acquisitions. The Wheatstone bridge is fully integrated on the probe. In the alternative circuit used to calibrate the DC response of the probe, the batteries are replaced by a AC symmetrical voltage generator and the filter is replaced by the input of a lock-in amplifier synchronized to the AC generator.}
\end{figure}

Two different electronic circuits are used: one dedicated to high signal-to-noise fluctuation measurements (see figure \ref{sketchElecFluct}) and the second one to accurate measurements of mean values. Two \SI{9}{\volt} batteries in series polarise the circuitry, and their common pole is grounded to the cryostat. Two similar resistors in series with the batteries allow  to tune the polarisation voltage of the Wheatstone bridge integrated on the probe. The typical polarisation of the cantilever is \SI{43.6}{mV} ($\approx$ \SI{90}{\micro \ampere}) and \SI{1}{\volt} ($\approx$ \SI{175}{\micro \ampere}) for the Pitot tube. The output signal is amplified directly on the top of the cryostat, using a low-noise AC preamplifier (EPC1-B), then anti-alias filtered by a KEMO 4th order filter. The acquisition is performed with a 18-bit multi-channel card (National Instrument 6289). The cut-off frequency $f_c$ of the filter is chosen to satisfy the Shannon criterion ($f_c < f_s/2$, with $f_s$ the sampling frequency). A  numerical low-pass filter at 800 Hz further reduces the bandwidth to discard frequencies altered by the $0.7 nV/\sqrt{Hz}$ noise floor, which is reached around 1 kHz and corresponds to the voltage noise of the preamplifier.
In this configuration, the frequencies below $\sim$10 mHz are rejected by the AC-preamplifier. That's why a dedicated DC electrical circuit is needed to measure the mean response of the probes. This is done replacing the batteries with a symmetrical \SI{10}{\hertz} AC source and performing lock-in detection (NF LI5640) on the pre-amplified output signal.
Although no signal distortion nor probe over-heating was found, as a precaution, the AC driving voltage is chosen to be equivalent to the one of the circuit with batteries.

\begin{figure}
\begin{center}
\includegraphics[width=7.5cm]{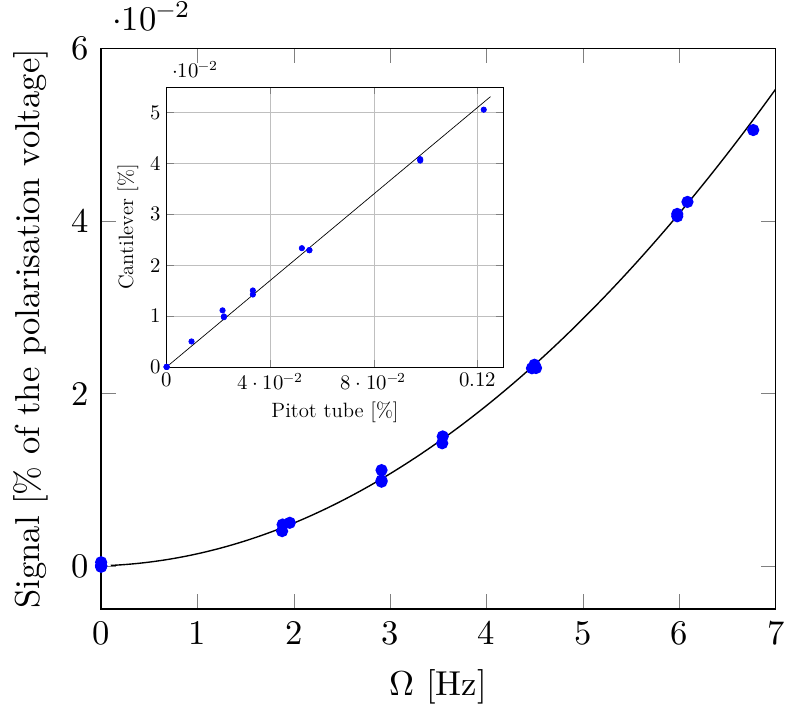}
\end{center}
\caption{\label{MeanResponse}Mean response of the cantilever probe versus the rotating velocity of the propeller, $\Omega$, in [Hz]. $\Omega$ can be considered as an image of the mean flow velocity at first order. The signal of the cantilever is quadratic versus velocity and linear versus signal of the Pitot tube (see insert) as illustrated by the parabolic and linear fits.}
\end{figure}

Calibration is performed in-situ using the mean response curves and a quadratic fit of the mean signal versus rotating velocity of the propeller (see Fig. \ref{MeanResponse}). %XXX
This response is fully consistent with the one obtained in air \cite{salortMEMS2}.
It is then possible to reconstruct the complete signal of the probes by combining the AC and the DC measurements. Surely, AC frequencies below 10 mHz are not fully recovered with this procedure but this has no consequence on the results of the present study

\paragraph*{Validation of the cantilever response:}

The cantilever beam is deflected by the hydrodynamic force imposed by the flow. As for the Pitot tube, this force is directly related to the dynamic pressure generated by the incoming flow. Above the superfluid transition, the typical Reynolds numbers based on the transverse size $l=32$ microns for the cantilever is the following:
\begin{equation}
Re_{canti} = {{\left< V \right>l} \over {\nu}} \approx 450 
\end{equation}
At such large Reynolds number, the dynamical pressure scales with the square of the velocity $p = \rho V^2 /2$ \cite{hoerner1965fluid}. The signal of the cantilever should then be quadratic with respect to the rotating velocity of the pump, which is actually the case (see figure \ref{MeanResponse}), and linear with respect to the Pitot tube signal, as confirmed by the insert.

\section{Results and discussions}

\label{results}

\subsection{Flow characterisation.}

Using the so-called Reynolds decomposition, velocity fluctuations are defined as:
 \begin{equation}
 v = V - \left< V \right>.
 \end{equation}
 
\begin{figure}
\begin{center}
\includegraphics[width=9cm]{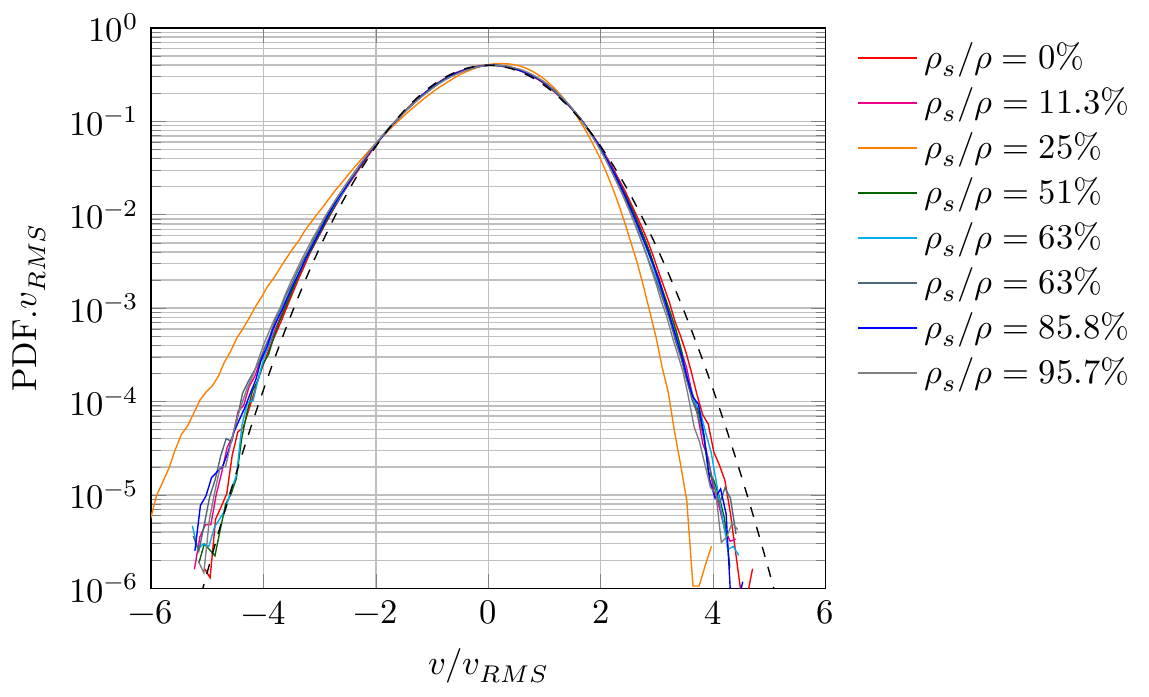}
\end{center}
\caption{\label{PDFVit}PDF obtained at different mean temperatures and nearly constant mean velocity ($\Omega \approx \SI{3.5}{\hertz}$). The dashed line corresponds to a a gaussian distribution.}
\end{figure}

Figure \ref{PDFVit} shows the probability density functions (PDF) of fluctuations in root mean square (noted RMS) units. 
All time series plotted are obtained for the same propeller rotation ($\Omega \approx \SI{3.5}{\hertz}$) and thus for nearly the same mean velocity.
Colors correspond to different temperatures except for the 1.85K temperature ($\rho_s / \rho = 63\%$), which has been achieved twice  and is represented using two different colors. Except the $\rho_s / \rho = 25\%$ time series (\SI{2.1}{\kelvin}) , all the PDFs remain close to a gaussian, with a small residual dissymmetry (skewness $v^3/({v_{RMS}}^{3/2})$  within [-0.16 ; -0.11]). This suggests that turbulence is not yet completely developed at 20 diameters downstream the disc. 
 We have no explanation for the odd behavior of \SI{2.1}{\kelvin} time series ; one could speculate on the appearance of a flow instability producing a recirculation or a corner-flow near the disc. Unfortunately, we discovered this odd behaviour too late to repeat the measurements. 

\begin{figure}
\begin{center}
\includegraphics[width=8.5cm]{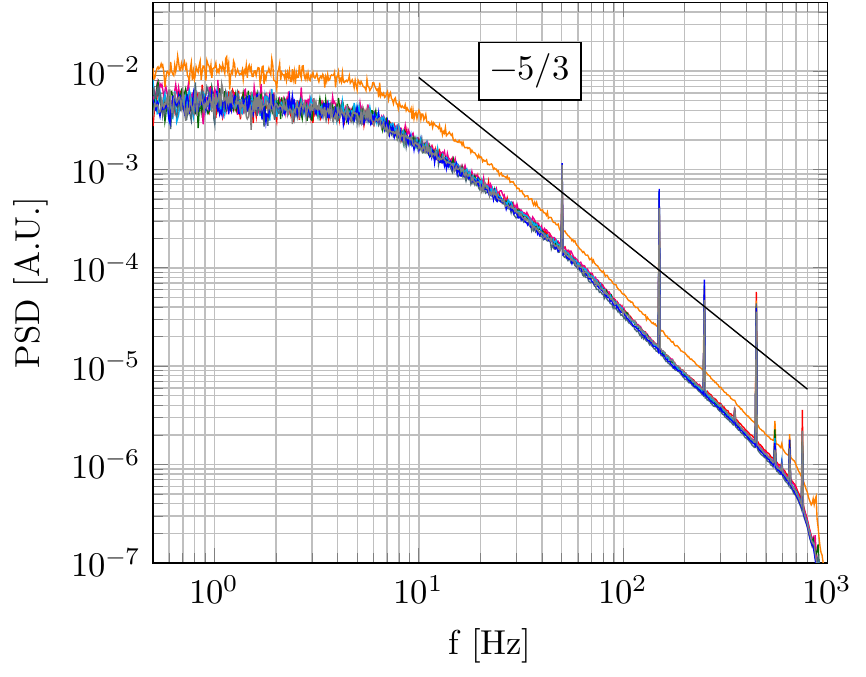}
\end{center}
\caption{Spectra of the velocity measured with the cantilever probe. The color code is the same as in figure \ref{PDFVit}. The cut-off at high frequency results from signal filtering.}
\label{spectra}
\end{figure}

Frequency spectra are presented in figure \ref{spectra}. As previously said, the time series are numerically filtered at \SI{800}{\hertz}, which explains the corresponding cut-off. To improve this signal-to-noise, a higher polarisation voltage would have been necessary. Unfortunately,  higher polarisations have proved to be potentially destructive for the fragile micro-machined electrical tracks of the probe.

At low frequency (typ. $f \lesssim  10 Hz$), the spectra evidence a characteristic plateau of one-dimensional velocity spectra. Above \SI{10}{\hertz} typically, the slope gets close to $-5/3$, which is characteristic for a fully developed turbulent cascade regime \cite{uriel1995turbulence}. This slope has been  reported in previous superfluid experiments in various very high Reynolds number flows, such as Von Karman cells \cite{Tabeling,Rousset:RSI2014}, wind tunnel \cite{RocheVortexSpectrum:EPL2007}, disc wake \cite{salortTOUPIE}, grid flow \cite{salortTSF} and jet \cite{Duri:RSI2015}. A closer analysis shows that the slope becomes slightly steeper than -5/3 in the second half of the resolved inertial range (roughly above 140 Hz).
This is consistent with an incomplete development of the turbulent cascade, 
and consistent with the observations of \cite{Johansson, Johansson2} in the wake behind a disc with a classical flow in conditions compatible with the present ones. A peak compatible with the vortex shedding frequency could have been expected around \SI{1}{\hertz} typically, which is not the case here. Two explanations are possible. First, in unconfined flow, the appearance of the peak is dependent on the radial position of the probe, in particular the peak can disappear at the center of the wake. 
Second, in some flows, 
the phenomena of vortex shedding are not present for specific ranges of Reynolds number compatible with the present one, as shown by \cite{Bearman} in the wake of cylinders. 

The spectrum associated with the \SI{2.1}{\kelvin} time series differ from the others, again. Its spectrum is more energetic, which is consistent with the appearance of a large scale flow instability in the tunnel, feeding more energy in the cascade. Considering that our main interest is not in this range of superfluid density ratio, we will not exploit this temperature in the following. Since no difference was found between the two times series independently recorded at 1.85K, only one will be displayed in the following \revision{figures}.

As a  test of data convergence, we examine third order statistics of velocity increments, which reveals the energy cascade process from large scales to small ones. 
The increments $\delta v$ of the longitudinal velocity $V$ in the $x$ direction parallel to the mean flow are defined as :
 \begin{equation}
\delta v = V(x+\delta x)-V(x),
 \end{equation}

\begin{figure}
\begin{center}
\includegraphics[width=8.5cm]{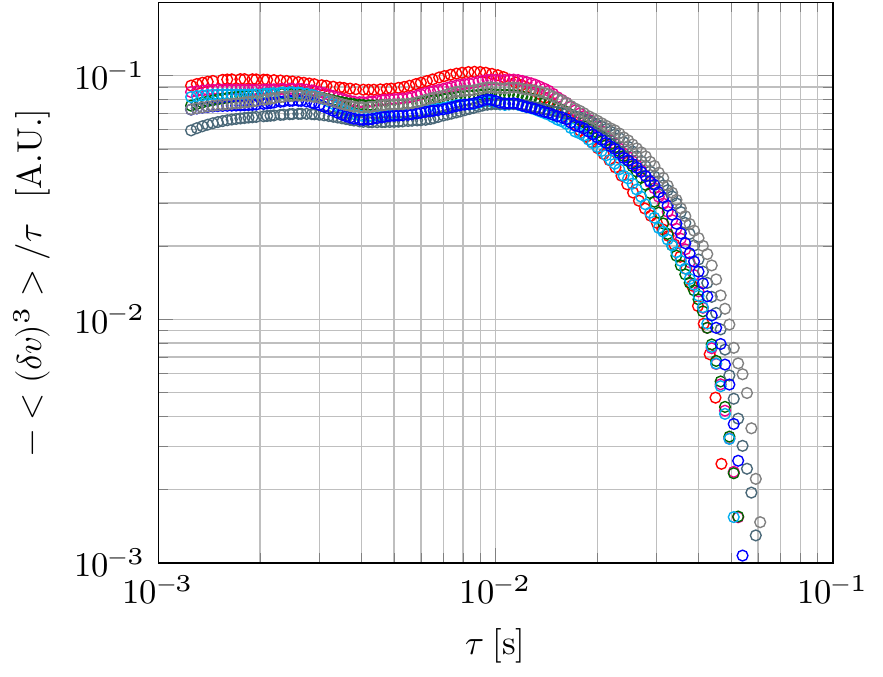}
\caption{\label{QuatreCinq}Third order structure function, same color code as in figure \ref{PDFVit}.}
\end{center}
\end{figure}

Taylor's frozen turbulence hypothesis is used to map the time domain, where the time series $V(t)$ are acquired, to the space domain $V(x)$, where the velocity increments are defined. This mapping is justified by the low turbulent intensity of the present flow, close to $7\%$. Using one of the datasets, we checked that use of the instantaneous Taylor hypothesis \cite{pinton1994correction} was not changing significantly the intermittency exponents (and only slightly accounting for the residual velocity skewness). The negligible influence on scaling exponents of this improved Taylor hypothesis was already pointed in the original paper \cite{pinton1994correction}.
 In practice, velocity increments will be directly estimated in the time domain as 
 \begin{equation}
 \delta v = V(t)-V(t+\tau)
  \end{equation}

\noindent  with $\tau=\delta x /  \left< V \right>$. The $4/5$ law of turbulence predicts the inertial-range scaling of the skewness of velocity increments  :
 \begin{equation}
 \left<  \delta v ^3 \right>  = - \frac{4}{5} \epsilon \cdot \delta x = - \frac{4}{5} \epsilon  \left< V \right>  \tau
 \end{equation}

\noindent where $\left< ... \right>$ denotes time averaging.
At the  Reynolds number of the present study ($R_\lambda \sim 400$), one does not expect a well defined plateau when plotting  $\left<  \delta v ^3 \right> / \tau$ versus $\tau$ due to finite Reynolds number correction. The 4/5 prefactor itself (not measurable in our experiment due to calibration uncertainty) is expected to be only approximatively reached (typ. within 10\%) in the middle of the inertial range (eg. see \cite{Qian:1999p3762,Antonia:2006p82} and reference within). With this in mind, one can still distinguish in Figure \ref{QuatreCinq} a clear leveling of this compensated third moment in the inertial range, which is consistent  
 with the literature (see \cite{Antonia,coscarella2017turbulent} for example) even if we do not resolve the small scales where $\left<  \delta v ^3 \right>$ is expected to decrease to 0.

\subsection{Determination of intermittency exponents.}

There exist several ways to quantify intermittency and this topic is still debated  (eg. see discussion in \cite{tsinober2013essence}). The motivation of the present work is not to obtain  \textit{absolute precision} in  coefficient characterizing intermittency but rather to obtain \textit{sensitivity} in the determination of these coefficients versus  temperature.  
This motivated the choice of a wake flow and led us to use the so-called ``extended self similarity'' (ESS) method \cite{BenziESS} to quantify intermittency through a set of scaling exponents $\zeta_p$ defined in the (extended) inertial range as :
 \begin{equation}
 \left< | \delta v | ^p \right> \sim {\left( \left< | \delta v | ^3 \right> \right)}^{\zeta_p}
 \label{EqRelative} 
 \end{equation} 

This method produces extended scaling ranges, which allows an accurate determination of the exponents ${\zeta_p}$. One  drawback of this method is the (small) difference between the ESS exponents ${\zeta_p}$ and the exponents ${\zeta_p^\prime}$ resulting from the ``genuine'' definition $ \left< \delta v ^p \right> \sim \delta x ^{\zeta_p^\prime}$. This drawback is a-priori not an issue here, since we focus on the relative \textit{variation} of exponents versus temperature. We will come back on this point in the conclusion section.

As a preliminary test of statistical convergence, we computed the histograms of  $| \delta v |^p$ up to $p=6$ and checked that their tails well converge to zero.
To determine the ESS exponents ${\zeta_p}$, we focus on their deviation from the exponents $p/3$ that would be expected in the absence of intermittency.
Thus, the \textit{intermittency corrections} $\mu_p =   {p / 3} - \zeta_p$ are directly fitted using a compensated log-log plot of  $\left<  | \delta v | ^p \right> \left< | \delta v | ^3 \right>^{-p/3}$ versus $\left< | \delta v | ^3 \right>$, or more precisely $-\mu_p$ is fitted as the slope of the affine function:
 \begin{equation}
\log \left({{  \left< |\delta v|^p \right>} \over { \left< |\delta v|^3 \right> ^{p/3}}}\right) =  -\mu_p  \cdot \log(\left< | \delta v |^3 \right>)+cst.
\end{equation}

\begin{figure}
\begin{center}
 \includegraphics[width=8.5cm]{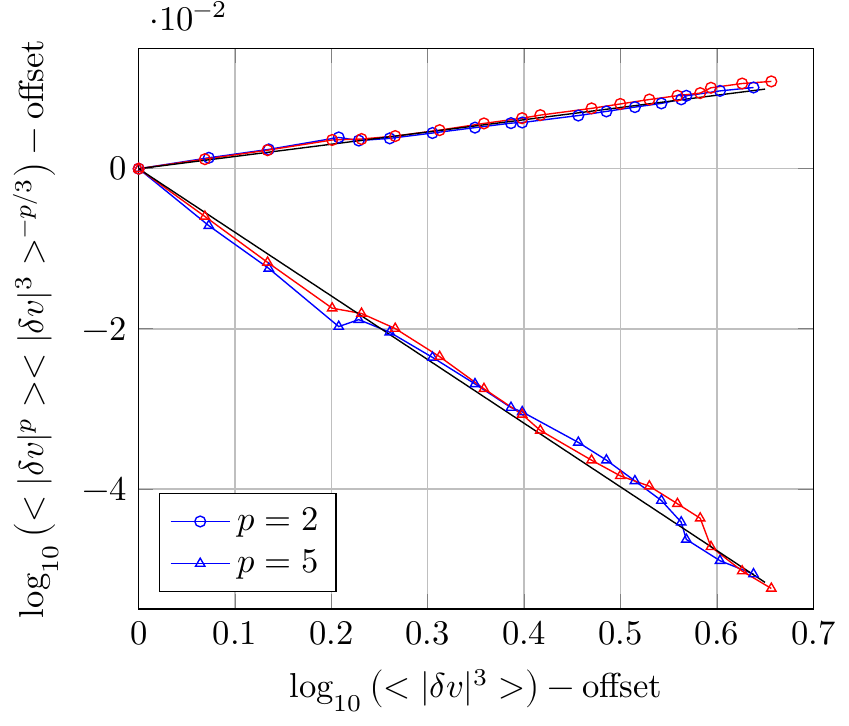}
\caption{\label{Fit} Example of the determination of the intermittency correction $\mu_p =   {p / 3} - \zeta_p$  (here at 2.32 K for the blue curves and 2.15 K for the red ones). In this  representation, the mean slope of each set of points is $-\mu_p$ . The time increments are windowed in the frequency range 20 - 140 Hz. Black lines are the fit of the \SI{2.32}{\kelvin} case.}
\end{center}
\end{figure}

This fit was performed for time increments chosen within 0.007-\SI{0.05}{\second} (i.e. 20-140 Hz), a range of increments which avoids the highest frequency part of the spectrum where the cascade is not fully developed. Although this range of increment is limited  to 0.84 decade, the good statistical converge of the data allows an accurate determination of a local exponent $\mu_p$, as illustrated by figure \ref{Fit}. This accurate determination should also be credited to the ESS method, which partly compensate for the absence of a pure scaling over the spectral range 20-140 Hz. As a check, a reduced range of time increments  (20-80 Hz) will also be used. The small steps visible for the $p=5$ datasets of Fig. \ref{Fit} are also present for the other orders and are interpreted as noise. In this representation, they don't alter significantly the slope determination, and therefore exponent determination. They would have been more detrimental if we were estimating exponents using the derivative ${d \log{ \left< |\delta v|^p \right>} } / {d \log(\left< | \delta v |^3 \right>) }$, and that's why we didn't use this alternative approach.

\begin{figure}
\begin{center}
\includegraphics[width=9cm]{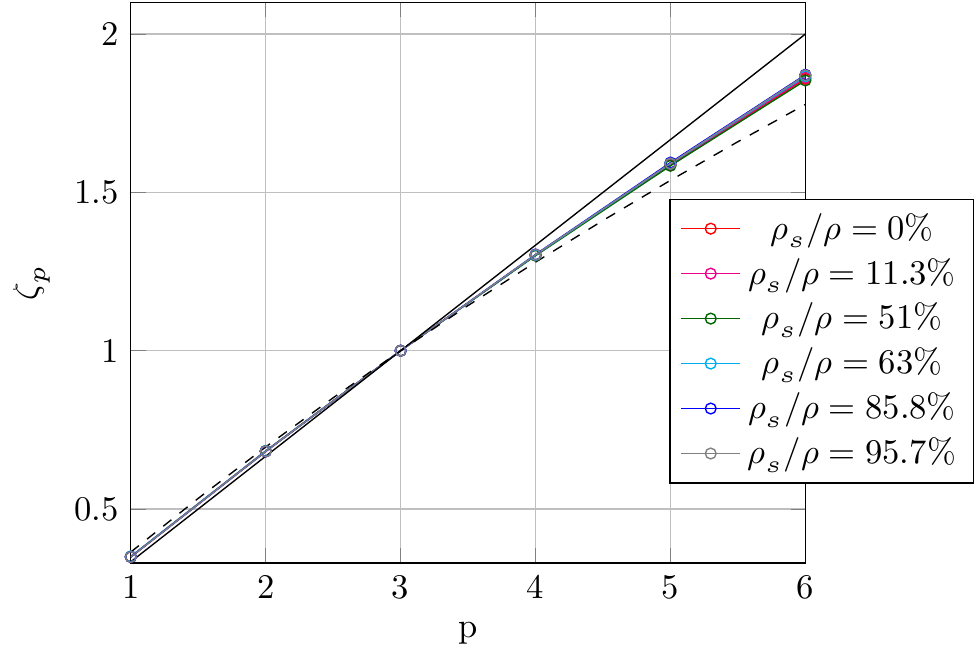}
\caption{\label{ExsposantPlot}Experimental exposant computed using the ESS method (coloured circles). The error bars corresponding to the fit uncertainty reported in Table \ref{Exsposant} are not plotted because there are smaller than the size of the circle symbols.
For comparison, are also plotted exponents $\zeta_p=p/3$ expected with no intermittency (black line), and those from the She-Lévêque model \cite{SheLeveque} ($\zeta_p^\prime = {p \over 9} + 2[1 - ({2 \over 3})^{p/3}]$), predicted using the standard  definition of exponents (dashed curve).}
\end{center}
\end{figure}

All the structure function exponents $\zeta_p$ and their fitting uncertainties are reported in Table \ref{Exsposant} and plotted in Fig. \ref{ExsposantPlot}. The error bars associated with the uncertainties are too small to worth plotting on Fig. \ref{ExsposantPlot} and later figures. %XXX
The exponents derived from Kolmogorov's 1941 self-similarity arguments (absence of intermittency, $\zeta_p=p/3$) and those from the She-Lévêque model \cite{SheLeveque} ($\zeta_p^\prime = {p \over 9} + 2[1 - ({2 \over 3})^{p/3}]$) are plotted for comparison. A direct quantitative comparison with the later model is delicate due to our use of the ESS method and the lack of isotropy and homogeneity of wake flows, but we can state that the flow presents the characteristics  features of intermittency (e.g. $\mu_2<0$ and $\mu_4 , \mu_5, \mu_6 >0$) and is quantitatively consistent with previous velocity fluctuations measurements done using a miniature Pitot tube in a perfectly homothetic confined wake geometry \cite{salortInter}. % XXX

The main result of this study is the following : up to uncertainties and over the full temperature range explored, intermittency is found independent from the superfluid fraction, including the intermediate temperature cases where a pronounced temperature dependence was reported in some numerical studies \cite{LvovInter,Shukla}.

\begin{center}
\begin{table}
\begin{tabular}{c | c  c  c  c  c}
$T$ [\SI{}{\kelvin}] & $\zeta_1$ $\pm 0.2\%$& $\zeta_2$ $\pm 0.1\%$& $\zeta_4$ $\pm 0.2\%$& $\zeta_5$ $\pm 0.5\%$& $\zeta_6$ $\pm 0.7\%$\\
 \hline
 \hline
2.32 & 0.349 & 0.682 & 1.302 & 1.585 & 1.86 \\
2.15 & 0.350 & 0.683 & 1.301 & 1.59 & 1.86 \\
%2.1 & &&&&\\
1.95 & 0.349 & 0.683 & 1.300 & 1.585 & 1.85 \\
1.85 & 0.348 & 0.681 & 1.303 & 1.59 & 1.86 \\
1.55 & 0.348 & 0.681 & 1.304 & 1.595 & 1.87 \\
1.28 & 0.348 & 0.682 & 1.303 & 1.59 & 1.87 \\
\end{tabular}
\caption{\label{Exsposant}Structure function exponents calculated with the ESS method in the 20-140 Hz range.}
\end{table}
\end{center}

\subsection{Comparison with previous studies}

\begin{figure}
\begin{center}
\includegraphics[width=9.0cm]{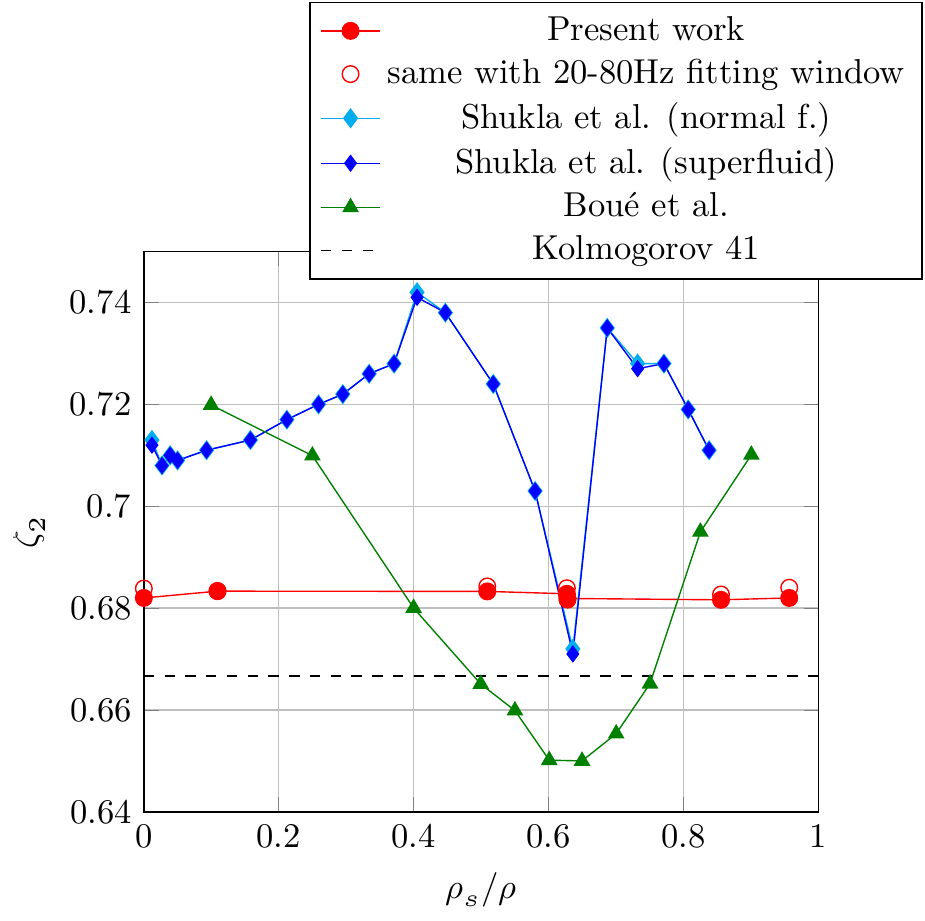}
\end{center}
\caption{\label{Exponent2}Exponents of the second order structure function as a function of the superfluid fraction. For explanation on open symbols, see text.}
\end{figure}

A preliminary comment is needed before comparing the exponent $\zeta_p$ from  experiment and numerics. Since the  anemometer is  sensing (one component of) the barycentric velocity $V= V_s \rho_s/\rho + V_n \rho_n/\rho$, the experimental exponents $\zeta_p$ are therefore characterizing this specific velocity.
In shell-model simulations, the normal fluid and superfluid velocity fields are modeled separately by discrete complex variables $u^n_m$ and $u^s_m$, one for each shell of index $m$ (wavelength). Exponents are therefore computed separately for each fluid component. Still, due to the strong coupling between the two fluids, they are nearly locked together in the inertial range  ($V_s \simeq V_n \simeq V$), which implies that the normal and superfluid exponents are similar. This is indeed the case in the numerics as illustrated in Fig.\ref{Exponent2} , in the supplemental materials of \cite{Shukla} (see the G1-G21 subsets, which are obtained using the fluid properties of He-II) and by the figure 1 from \cite{LvovInter} which shows similar normal and superfluid structure functions  in the inertial range, implying similar intermittency exponents. It is therefore fair to compare the exponents from the experiment and numerics. Surely, this would no longer be straightforward if we were studying small-scale intermittency, and not inertial-range intermittency.

To summarize the existing results,  experiments  (\cite{Tabeling,salortInter} and present study) and simulations (\cite{salortInter,LvovInter,Shukla}) did not reveal any difference of intermittency between classical turbulence and quantum turbulence in both temperature limits: high  ($\rho_s / \rho \ll 1$) and low (but finite) temperature ($0.04 \lesssim \rho_n / \rho \ll 1$). In the intermediate temperature range, the present experiment exhibits no difference between the classical and quantum cases up to an excellent resolution, in contradiction with shell-model simulations predicting  significant enhancement  \cite{LvovInter} or reduction\cite{Shukla}.

To illustrate quantitatively the disagreement between our experiment and both shell simulations, we plot in fig. \ref{Exponent2} the second order exponent $\zeta_2$ from these three studies. The values in classical (Navier-Stokes) limit $\zeta_2^{NS} = \zeta_2(\rho_s =0)$ differ between the shell models (0.72) and our experiment results ($\simeq 0.68$) but this should not be considered as an issue. Indeed, the absolute value of $\zeta_2^{NS}$ results from an arbitrary choice of model parameters in shell simulation (as recalled in \cite{LvovInter}) and it is biased by use of the ESS method in experiments, as already explained, and possibly by residual non-homogeneity and anisotropy of wake flows. To check if the 20-140\, Hz windowing of the time increments as a significant impact of the fitted exponents, the reduced window 20-80\,Hz was also used. The open symbols in fig. \ref{Exponent2} show that the impact is limited.
 The most striking features of this figure are the difference in temperature dependence between the three studies. Interestingly, the exponents  $\zeta_2$ obtained in the simulations by Shukla et al. \cite{Shukla}  both exceed and 
fall short of their classical limit $\zeta_2^{NS}$, which could be interpreted respectively as an intermittency enhancement and reduction. In Boué et al.  simulations\cite{LvovInter}, the exponents $\zeta_2$ have a minimum below Kolmogorov 1941 value $\zeta_2=2/3$, which corresponds itself to an absence of intermittency. The authors interpretation of an "enhancement" of intermittency (instead of the apparent cancellation) is based on higher order exponents.

\begin{figure}
\begin{center}
\includegraphics[width=8.5cm]{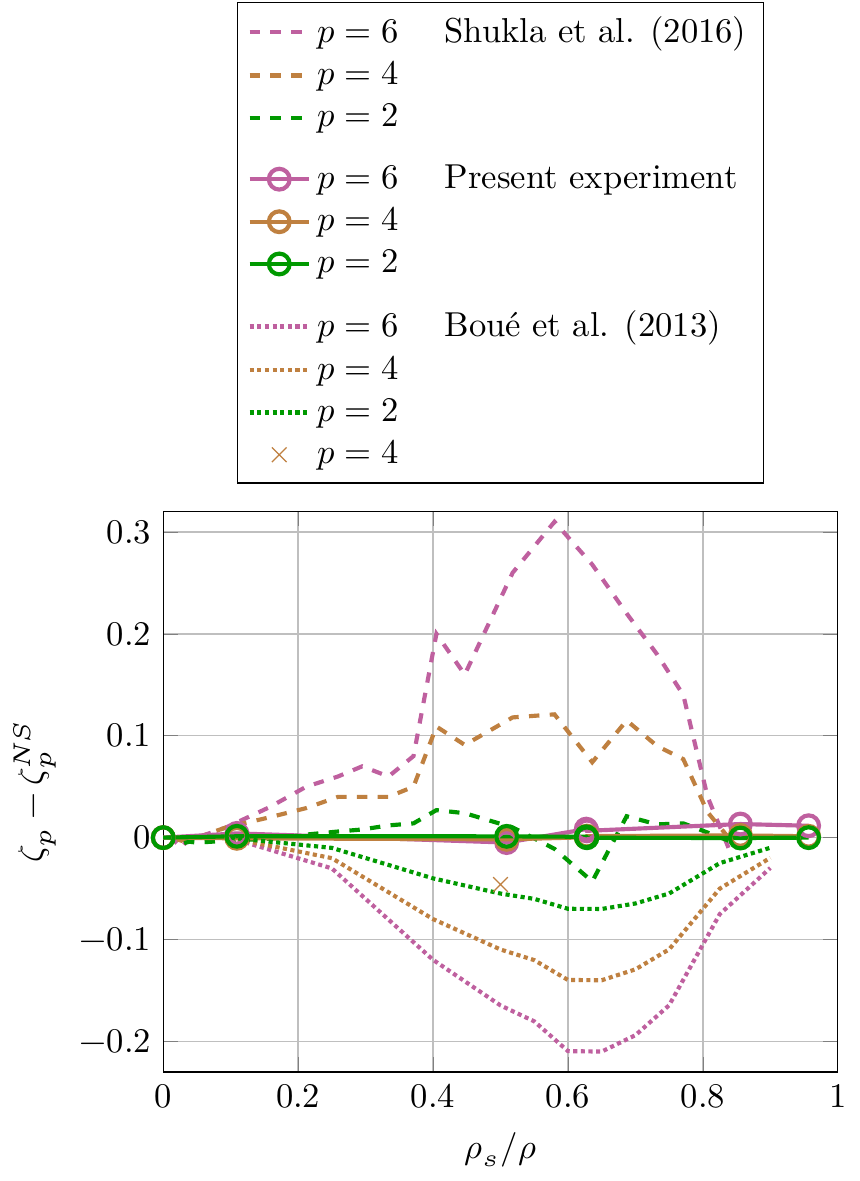}
\end{center}
\caption{\label{ExponentP-NS}Superfluid correction of the intermittency exponents. Note that the dotted line for orders $p=4$ and $p=6$ have been calculated from an analytical formula provided in the original paper.}
\end{figure}

To focus on possible superfluid effect on the intermittency, we consider now the relative exponents: 
 \begin{equation}
\zeta_p - \zeta_p^{NS} =  \zeta_p - \zeta_p (\rho_s = 0) \simeq \zeta_p - \zeta_p (\rho_s \rightarrow 0)
 \end{equation} 

\noindent which can be seen as the superfluid correction to the classical exponent. Since all studies agree that the classical exponents $\zeta_p^{NS}$ are recovered in the $\rho_s/\rho \rightarrow 0$ limit, this definition allows to single out only superfluid effects.

Figure \ref{ExponentP-NS} represents this superfluid intermittency correction on exponents for $p=2,4,6$. To put numbers on Eq.9, values from Shukla et al. simulations are taken from the supplemental materials of their article \cite{Shukla}. 
Boué et al.  paper \cite{LvovInter} provides one value  $\zeta_4 \simeq 1.21$ for $\rho_s/\rho=0.5$ and $\zeta_4^{NS} = 1.256$ (see cross in Fig. \ref{ExponentP-NS}), a plot of $\zeta_2$ and a relation for  $\zeta_p$ versus $\zeta_p^{NS}$ and $\zeta_2$  ``in good agreement with the observed values'' (with our notations, they found {$\zeta_p - \zeta_p^{NS} = p \left( \zeta_2 - \zeta_2^{NS} \right) / 2$ })  which allowed us to estimate the complete Figure \ref{ExponentP-NS}. Like for the previous figure, the differences between the three studies are striking : no superfluid effect is found in the present experiment, while strong opposite effects reported in the shell simulations. This is the central experimental result of this study.

\section{Concluding remarks}
\label{Conclusion}

We measured intermittency in the upper inertial range of a turbulent cascade of superfluid $^4$He, with a special attention for the intermediate temperatures where none of the two fluids components of He-II can be neglected. In this range of temperature, no other experimental data were published and two published simulations are giving contradictory results  : Boué et al. predicting an excess of intermittency  \cite{LvovInter} and Shukla et al. a deficit of it \cite{Shukla}.
Our measurements disagree with both simulations:  we don't detect any temperature dependence of scaling exponents (with better than $\pm0.7$\% precision up to 6th order) when temperature is varied between the Navier-Stokes limit ($\rho_s =0$ for $T=2.32K$) down to 1.28K, where 96\% of He-II is superfluid. Our results also contradicts a LES simulation claiming an enhancement of intermittency near 1.6K \cite{Bakhtaoui:2014}.

Understanding the reason for the disagreements between the shell-model simulations \cite{LvovInter,Shukla} is beyond the scope of this paper. As acknowledged by the authors of these numerics, it is not surprising that shell-model simulations recover the classical intermittency exponents in the low and high temperature limits. Indeed, in these limits, the fluid with the largest density fully controls the dynamics without being significantly disturbed by the low-density one (which follows the former, due to strong coupling). Thus, one recover a one-fluid dynamical system with an inter-shell coupling term $NL[u_m^{n,s}]$ and numerical coefficients ``$a,b,c$''  which had been specially tuned to recover the classical exponents. The disagreement between both simulations (not to mention the experiment) at intermediate temperatures question the ability of the traditional inter-shell-coupling model to capture the intermittent corrections in presence of mutual coupling between superfluid and normal-fluid, at least for the mutual coupling model implemented in both simulations. To go beyond, a systematic study of the sensitivity of scaling exponents versus shell-model parameters could be interesting. Further studies, in particular high-resolution DNS simulations will probably be of great help. Efforts in this direction are underway by different groups.

We now come back to the comparison between the shell-model simulations and the experiments. The simulations provide the \textit{absolute} scaling exponents $\zeta_p^\prime$ defined as $ \left< | \delta u_m | ^p \right> \sim {k_m}^{-\zeta_p^\prime}$ ($k_m$ is the wavevector of the $m^{th}$ shell), which is the shell-model version of
the definition $ \left<| \delta v | ^p \right> \sim \delta x ^{\zeta_p^\prime}$. The ESS method used for the experiment produces \textit{relative} scaling exponents $\zeta_p$  (see  Eq.\ref{EqRelative}) defined with respect to the third moment, which is expected to scale linearly with $\delta x$ in the inertial range of homogenous isotropic turbulence. It has been noticed that (inertial range) absolute exponents $\zeta_p^\prime$ determined from shell-model simulations can be sensitive to the dissipative processes occuring at small scales, while relative exponents $\zeta_p \sim \zeta_p^\prime / \zeta_3^\prime$ are not \cite{LevequeShe1995}. A-priori, this could have explained the observed discrepancy between experiment and shell-model, but it is not the case here, as can be seen in two ways. First, if the absolute exponents of $\zeta_2^\prime$ in the present study had the 10\%  temperature dependence found in the simulations, the spectra of fig. \ref{spectra}  would not overlap as well. Second, when the absolute exponents reported in the shell-model simulations \cite{LvovInter,Shukla} are normalized by the third order exponent, we find that $\zeta_p^\prime/\zeta_3^\prime$ still have a significant temperature dependence. Thus, the difference of definition of scaling exponents cannot explain the qualitative difference between these simulations and the experiment.

On the experimental side, it would be interesting to extend the result to purely homogeneous and isotropic conditions. The use of a grid to generate turbulence would have produced a more ``ideal'' flow, but also smaller length scales and a smaller level of velocity fluctuations, resulting in a significantly lower range of resolved scales given to finite resolution and sensitivity of probes. New probe and flow designs would therefore be required to go in this direction. Regarding present results, we only explored the inertial range over nearly 1 decade of scales (the largest ones), and we cannot exclude that a different picture may emerge at  smaller scales. In particular, it would be interesting to explore length scales closer to the mesoscale ``grey'' zone, where strong difference in dynamics between the superfluid and normal fluid are expected to appear and a partial randomization (or equipartition) of the superfluid excitations has been predicted \cite{Salort:EPL2011}. 

\bigskip

\noindent \textbf{ACKNOWLEDGMENTS}

We thank G. Garde for the mechanical design and realization of the experimental apparatus, E. Verloop for the pumping group electrical control system and G. Bres for the specific liquid helium level electronics. We are also grateful to F. Chillà and B. Castaing for their participation in the design of the cantilever, to Y. Gagne, E. Lévêque and T. Dombre for sharing their insights on intermittency and shell models, B. Hébral for his feed-back and S. Hatzikiriakos -as the associate editor of Physics of Fluids- for finding good referees, also thanked here. We acknowledge financial support from EC Euhit project (WP21), which enabled the development of probes, financial support from the ANR SHREK for the pumping group, and support from the ANES.

\bibliographystyle{unsrt}
%\bibliography{References.bib}

\begin{thebibliography}{10}

\bibitem{DonnellyLivreVortices}
R.~J. Donnelly.
\newblock {\em Quantized Vortices in Helium-{II}}.
\newblock Cambridge Studies in Low Temperature Physics. Cambridge University
  Press, Cambridge, 1991.

\bibitem{VanSciverLivre2012}
S.W. Van~Sciver.
\newblock {\em Helium Cryogenics}.
\newblock International Cryogenics Monograph Series. Springer, 2012.

\bibitem{BarenghiSkrbekSreenivasan_IntroPNAS2014}
C.~F. Barenghi, L.~Skrbek, and K.~R. Sreenivasan.
\newblock Introduction to quantum turbulence.
\newblock {\em Proceedings of the National Academy of Sciences}, 111(Supplement
  1):4647--4652, 2014.

\bibitem{spectra:PNAS2014}
C.~F. Barenghi, V.~S. L'vov, and P.-E. Roche.
\newblock Experimental, numerical, and analytical velocity spectra in turbulent
  quantum fluid.
\newblock {\em PNAS}, 111((Supplement 1)):4683--4690, 2014.

\bibitem{salortTOUPIE}
J.~Salort and \textit{et al}.
\newblock Energy cascade and the four-fifths law in superfluid turbulence.
\newblock {\em Eur. Phys. Lett.}, 97:34006, 2012.

\bibitem{uriel1995turbulence}
Uriel Frisch.
\newblock {\em Turbulence: The Legacy of A. N. Kolmogorov}.
\newblock Cambridge University Press, 1995.

\bibitem{Sreenivasan:1997p84}
K.~R. Sreenivasan and R.~A. Antonia.
\newblock The phenomenology of small-scale turbulence.
\newblock {\em Annual Review of Fluid Mechanics}, 29:435, Jan 1997.

\bibitem{tsinober2013essence}
A.~Tsinober.
\newblock {\em The essence of turbulence as a physical phenomenon: with
  emphasis on issues of paradigmatic nature}.
\newblock Springer Science \& Business Media, 2013.

\bibitem{benzi2015homogeneous}
R.~Benzi and L.~Biferale.
\newblock Homogeneous and isotropic turbulence: A short survey on recent
  developments.
\newblock {\em Journal of Statistical Physics}, 161(6):1351--1365, 2015.

\bibitem{Tabeling}
J.~Maurer and P.~Tabeling.
\newblock Local investigation of superfluid turbulence.
\newblock {\em Europhys. Lett.}, 43:29, 1998.

\bibitem{salortInter}
J.~Salort, B.~Chabaud, E.~L\'ev\^eque, and P.E. Roche.
\newblock Investigation of intermittency in superfluid turbulence.
\newblock {\em Jour. Phys. : Conf. Series}, 318, 2011.

\bibitem{LvovInter}
L.~Boué and \textit{et al}.
\newblock Enhancement of intermittency in superfluid turbulence.
\newblock {\em Phys. Rev. Lett.}, 110:014502, 2013.

\bibitem{biferale2003shell}
L.~Biferale.
\newblock Shell models of energy cascade in turbulence.
\newblock {\em Annual review of fluid mechanics}, 35(1):441--468, 2003.

\bibitem{Shukla}
V.~Shukla and R.~Pandit.
\newblock Multiscaling in superfluid turbulence: A shell-model study.
\newblock {\em Phys. Rev. E}, 94:043101, 2016.

\bibitem{Bakhtaoui:2014}
M.~{Bakhtaoui} and L.~{Merahi}.
\newblock {Analysis of the Energy Budget in Quantum Turbulence: HVBK Model}.
\newblock {\em Journal of Low Temperature Physics}, 178:129--141, February
  2015.

\bibitem{Krstulovic:PRE2016}
G.~Krstulovic.
\newblock Grid superfluid turbulence and intermittency at very low temperature.
\newblock {\em Phys. Rev. E}, 93:063104, Jun 2016.

\bibitem{Rusaouen:parietalEPL2017}
E.~Rusaouen, B.~Rousset, and P.-E. Roche.
\newblock Detection of vortex coherent structures in superfluid turbulence.
\newblock {\em EPL}, 118:14005, 2017.

\bibitem{Kahal}
Kahalerras, H and Malecot, Y and Gagne, Y and Castaing, B
\newblock Intermittency and Reynolds number.
\newblock {\em Physics of Fluids},10:910, 1998.

\bibitem{Carmody1}
T.~Carmody.
\newblock Establishment of the wake behind a disk.
\newblock {\em Jour. Bas. Eng.}, 86:869 1964.

\bibitem{Cannon}
S.~Cannon, F.~Champagne, and A.~Glezer.
\newblock Observations of large-scale structures in wakes behind axisymmetric bodies.
\newblock {\em Exp. Fluid.}, 14:447, 1993.

\bibitem{Johansson}
P.~B.~V. Johansson, W.~K. George, and S.~H. Woodward.
\newblock Proper orthogonal decomposition of an axisymmetric turbulent wake behind a disk.
\newblock {\em Phys. Fluid.}, 14:2508, 2002.

\bibitem{Johansson2}
P.~B.~V. Johansson, S.~H. Woodward, and W.~K. George.
\newblock The far downstream evolution of the high-reynolds-number axisymmetric wake behind a disk. part 1. single-point statistics.
\newblock {\em J. Fluid Mech.}, 555:363, 2006.

\bibitem{mehta_bradshaw_1979}
Mehta, R.D. and Bradshaw, P.
\newblock Design rules for small low speed wind tunnels.
\newblock {\em The Aeronautical Journal (1968)}, 83:443--453, 1979.

\bibitem{salortMEMS1}
J.~Salort, P.E. Roche, and A.~Monfardini.
\newblock Cantilever anemometer based on a superconducting micro-resonator: Application to superfluid turbulence.
\newblock {\em Rev. Sci. Instr.}, 83:125002, 2012.

\bibitem{salortMEMS2}
J.~Salort and \textit{et al}.
\newblock Joint temperature and velocity local sensor for turbulent flows.
\newblock {\em subm. to Rev. Sci. Instr.}, 2017.

\bibitem{Sader_JApplPhys98}
John~Elie Sader.
\newblock Frequency response of cantilever beams immersed in viscous fluids
  with applications to the atomic force microscope.
\newblock {\em Journal of Applied Physics}, 84:64--76, 1998.

\bibitem{salortTSF}
J.~Salort and \textit{et al}.
\newblock Turbulent velocity spectra in superfluid flows.
\newblock {\em Phys. Fluids}, 22:125102, 2010.

\bibitem{Roche2fluidCascade:EPL2009}
P.-E. Roche, C.F. Barenghi, and E.~Leveque.
\newblock Quantum turbulence at finite temperature: The two-fluids cascade.
\newblock {\em EPL}, 87(5):54006, 2009.

\bibitem{hoerner1965fluid}
Sighard~F Hoerner.
\newblock {\em Fluid-dynamic drag: Practical Information on Aerodynamic Drag
  and Hydrodynamic Resistence}.
\newblock Sighard F. Hoerner, 1965.

\bibitem{Rousset:RSI2014}
B.~{Rousset}, P.~Bonnay, P.~Diribarne, A.~Girard, J.M. Poncet, E.~Herbert,
  J.~Salort, C~Baudet, B~Castaing, L.~Chevillard, F.~Daviaud, B.~Dubrulle,
  Y.~Gagne, M.~Gibert, B.~H{\'e}bral, T.~Lehner, P.-E. Roche, B.~Saint-Michel,
  and M~Bon~Mardion.
\newblock Superfluid high reynolds von k{\'a}rm{\'a}n experiment.
\newblock {\em Rev. Sci. Instrum.}, 85:103908, 2014.

\bibitem{RocheVortexSpectrum:EPL2007}
P.-E. Roche, P.~Diribarne, T.~Didelot, O.~Fran{\c c}ais, L.~Rousseau, and
  H.~Willaime.
\newblock Vortex density spectrum of quantum turbulence.
\newblock {\em EPL}, 77:66002, 2007.

\bibitem{Duri:RSI2015}
D.~Dur{\`\i}, C.~Baudet, J.-P. Moro, P.-E. Roche, and P.~Diribarne.
\newblock Hot-wire anemometry for superfluid turbulent coflows.
\newblock {\em Review of Scientific Instruments}, 86(2):025007, 2015.

\bibitem{Bearman}
P.W. Bearman.
\newblock On vortex shedding from a circular cylinder in the critical reynolds number regime.
\newblock {\em J. Fluid Mech.}, 37:577, 1969.

\bibitem{pinton1994correction}
Pinton, J-F and Labb{\'e}, R
\newblock Correction to the Taylor hypothesis in swirling flows.
\newblock {\em Journal de Physique II}, 4:1461--1468, 1994.

\bibitem{Antonia}
R.A. Antonia, T.~Zhou, and J.P. Romano.
\newblock Small-scale turbulence characteristics of two-dimensional bluff body wakes.
\newblock {\em J. Fluid Mech.}, 459:67, 2002.

\bibitem{Qian:1999p3762}
J Qian
\newblock Slow decay of the finite Reynolds number effect of turbulence  wakes
\newblock {\em Phys. Rev. E}, 60:3409--3412, 1999.

\bibitem{Antonia:2006p82}
R. A Antonia and P Burattini
\newblock Approach to the 4/5 law in homogeneous isotropic turbulence
\newblock {\em J. Fluid Mech.}, 550:175, 2006.

\bibitem{coscarella2017turbulent}
Coscarella, F and Servidio, S and Ferraro, D and Carbone, V and Gaudio, R
\newblock Turbulent energy dissipation rate in a tilting flume with a highly rough bed.
\newblock {\em Phys. Fluids}, 29:085101, 2017.
	
\bibitem{BenziESS}
R.~Benzi and \textit{et al}.
\newblock Extended self-similarity in turbulent flows.
\newblock {\em Phys. Rev. E}, 48:R29, 1993.

\bibitem{SheLeveque}
Z.-S. She and E.~Lévêque.
\newblock Universal scaling laws in fully developed turbulence.
\newblock {\em Phys. Rev. Lett.}, 72:336, 1994.

\bibitem{LevequeShe1995}
Emmanuel Leveque and Zhen-Su She.
\newblock Viscous effects on inertial range scalings in a dynamical model of
  turbulence.
\newblock {\em Phys. Rev. Lett.}, 75:2690--2693, Oct 1995.

\bibitem{Salort:EPL2011}
J.~Salort, P.-E. Roche, and E.~L{\'e}v{\^e}que.
\newblock Mesoscale equipartition of kinetic energy in quantum turbulence.
\newblock {\em EPL}, 94:24001, 2011.

\end{thebibliography}

\end{document}